\pdfoutput=1

\documentclass[referee]{raa}            % referee version: for submission
\usepackage{graphicx,times}             %for PS/EPS graphics inclusion, new
\usepackage{natbib}
\usepackage{amssymb,amsmath}
\usepackage{xcolor}
\usepackage{soul}

\bibpunct{(}{)}{;}{a}{}{,}

\def\cl{{\cal L}}
\def\ce{{\cal E}}
\def\bc{\color{blue}}

\usepackage[pagebackref=true]{hyperref}
\begin{document}

  \title{Charged Particles Capture Cross-Section by a Weakly Charged Schwarzschild Black Hole}

   \volnopage{Vol.0 (20xx) No.0, 000--000}      %%preserved for Editor. DOn't remove!
   \setcounter{page}{1}          %%starting page, preserved for Editor. DOn't remove!

   \author{A. M. Al Zahrani, 
      \inst{1}
   \and A. Al-Jama
      \inst{2}
   }

   \institute{Physics Department, King Fahd University of Petroleum and Minerals, Dhahran 31261, Saudi Arabia; {\it amz@kfupm.edu.sa} \inst{1}; {\it ahmad.aljama1905@gmail.com} \inst{2}\\
\vs\no
   {\small Received 20xx month day; accepted 20xx month day}}

\abstract{ We study the capture cross-section of charged particles by a weakly charged Schwarzschild black hole. The dependence of the maximum impact parameter for capture on the particle's energy is investigated numerically for different values of the electromagnetic coupling strength between the particle and the black hole. The capture cross-section is then calculated. We show that the capture cross-section is independent of the electromagnetic coupling for ultra-relativistic particles. The astrophysical implications of our results are discussed.   
}
   \authorrunning{A. M. Al Zaharni \& A. Al-Jamaa}            %author_head in even pages
   \titlerunning{Capture Cross-Section by Weakly Charged Black Hole}  % title_head in odd pages

   \maketitle
%% The author head (on even pages) and the title head (on odd pages) will be
%% automatically extracted from \author{} and \title{}. Whenever the title is too long,
%% you will be asked to supply a shorter one by inserting either \authorrunning{} or
%% \titlerunning{} before \maketitle. Anyway, you can specify your own heads.
%%
%%
%% Note: In the following text body of your manuscript, please note several differences from
%%       other major journals:
%% (1) \subsection{Please Capitalize the First Letter of Each Notional Word in Subsection Title}
%% (2) Please Capitalize the First Letter of Each Notional Word in all tables' captions

%
%________________________________________________ sections below
%
\section{Introduction}           %% first-level sections will be auto-capitalized
\label{sect:intro}

\noindent Studying the capture cross-section of black holes is central to understand how matter interacts with them. It helps us understand the process of matter accretion by a black hole which in turn determines how its mass, angular momentum and charge evolve. It can also help us understand the environment near black holes. Moreover, scrutinizing capture cross-section can be used to test theories of gravity in strong gravitational fields.
  
Astrophysicists generally assume black holes are electrically neutral. This is because they would quickly attract oppositely charged matter to balance out any access charge. However, there are compelling reasons why weakly charged black holes might exist as discussed in~\cite{Z1,Z2,Zaj1,Zaj2,Zaj3,Car} and the references therein. The differences in how a black hole accretes electrons and protons within its plasma environment, influenced by radiation, could render it charged. Also, the spin of a black hole in the presence of a magnetic field can induce the accretion of charged particles. In fact, using the EHT observations, it was inferred that Sgr A* and M87 can be charged~\cite{GA,EHT}. The black hole's charge is weak in the sense that it has no tangible effect on spacetime, but its effect on charged particle dynamics is prominent.

\noindent There are numerous astrophysical scenarios wherein charged particles are drawn into black holes. Stars within the Roche limits near black holes often contribute matter through tidal interactions. Additionally, stars emit streams of charged particles as stellar winds. Highly energetic charged particles, resulting from supernovae, gamma-ray bursts, and bipolar jets from compact objects, frequently find their way into the vicinity of black holes. These processes collectively enrich the environment around black holes with a significant population of charged particles.

\noindent The concept of capture cross-sections has been explored extensively for various black hole types.  Foundational treatment which examine photon and neutral particle capture by Schwarzschild black holes was given in several monographs, such as~\cite{FZ}. Further work addressed capture cross-sections of charged and neutral particles by Kerr-Newman black holes, including the implications for black hole spin and charge evolution~\cite{Youn}. Capture by Reissner-Nordström black holes was also investigated~\cite{Zak}. In the context of higher-dimensional black holes, studies have focused on calculating photon critical impact parameters for Schwarzschild-Tangherlini black holes~\cite{CoFr, Tsu, SiGh, Buq}. The capture cross-section for massive particles was determined in~\cite{Ahm}.  Additionally, research extends to particle capture in Myers-Perry rotating spacetime which describes rotating black holes in five-dimensions~\cite{GoFr}. Moreover, wave capture cross-sections have been studied for various black hole configurations (see~\cite{Ana} and the references within).

\noindent In this research, we examine the capture cross-section of charged particles by a weakly charged Schwarzschild black hole and discuss the astrophysical consequences of our findings. The paper is organized as follows: In Sec.~\ref{S2}, we review the dynamics of charged particles in the background of a weakly charged black hole. We then review the capture cross-section of neutral particles in Sec.~\ref{S3}. The capture cross-section of charged particles is calculated for different coupling strengths and particle energies in Secs.~\ref{S4}. Finally, we summarize our main findings and discuss their astrophysical consequences in Sec~\ref{S5}. We use the sign conventions adopted in~\cite{MTW} and geometrized units where $c=G=k=1$, where $k$ is the electrostatic constant.

\section{Charged Particles near a Weakly Charged Schwarzschild Black Hole}\label{S2}

Here, we review the dynamics of charged particles near a weakly charged black hole. The spacetime geometry around a black hole of mass $M$ and charge $Q$ is described by the Schwarzschild Reissner-Nordstr\"on metric which reads~\cite{MTW}
\begin{equation}
ds^2=-hdt^2+h^{-1}dr^2+r^2d\theta^2+r^2\sin^2{\theta}d\phi^2,
\end{equation}
where $h=1-r_S/r+Q^2/r^2$ and $r_S=2M$ is the Schwarzschild radius. The electromagnetic 4-potential is 
\begin{equation}
A_\mu=-\frac{Q}{r} \delta_\mu^0.
\end{equation}
However, when the charge is weak we can ignore the curvature due to it and use the Schwarzschild metric, which reads~\cite{MTW}
\begin{equation}
ds^2=-fdt^2+f^{-1}dr^2+r^2d\theta^2+r^2\sin^2{\theta}d\phi^2,
\end{equation}
where $f=1-r_S/r$ and $r_S=2M$ is the Schwarzschild radius. This weak charge approximation is valid unless the charge creates curvature comparable to that due to the black hole's mass. This happens when
\begin{equation}
Q^2 \sim M^2.
\end{equation}
In conventional units, the weak charge approximation fails when
\begin{equation}
Q \sim \frac{G^{1/2}M}{k^{1/2}} \sim 10^{20}\:\frac{M}{\:M_\odot}\:\text{coloumbs}.
\end{equation}
This charge is way greater than the greatest estimated change on any black hole. Although the black hole charge is tiny, its effect on charged particles dynamics is profound because it is multiplied by the charge-to-mass ratio of these particles ($\sim10^{21}\: {\text m}^{-1}$ for electrons and $\sim10^{18}\: {\text m}^{-1}$ for protons).

The Lagrangian describing a charged particle of charge $q$ and mass $m$ in a spacetime described by a metric $g_{\mu\nu}$ and an electromagnetic field produced by a 4-potential $A^\mu$ reads \cite{Cha}
\begin{equation}
L=\frac{1}{2}mg_{\mu\nu}u^\mu u^\nu+qu^\mu A_\mu,
\end{equation}
where $u^\mu\equiv dx^\mu/d\tau$ is the particle’s 4-velocity and $\tau$ is its proper time. In our case, the Lagrangian becomes
\begin{eqnarray}
 L &=&\frac{1}{2} m\left[-f\left(\frac{dt}{d\tau}\right)^2+f^{-1}\left(\frac{dr}{d\tau}\right)^2+r^2\left(\frac{d\theta}{d\tau}\right)^2+r^2\sin^2{\theta}\left(\frac{d\phi}{d\tau}\right)^2 \right] \nonumber \\
&& -qQ\frac{dt}{d\tau}.
\end{eqnarray}
This Lagrangian is cyclic in $t$ and $\phi$, which means that the particle’s energy and azimuthal angular momentum are constants of motion. The specific energy and azimuthal angular momentum are, respectively, given by
\begin{eqnarray}
{\cal E}&=&-\frac{1}{m}\frac{\partial L}{\partial\left(\frac{dt}{d\tau}\right)}=f\frac{dt}{d\tau}+\frac{qQ}{mr}, \\
\ell&=&\frac{1}{m}\frac{\partial L}{\partial\left(\frac{d\phi}{d\tau}\right)}=r^2\sin^2{\theta}\frac{d\phi}{d\tau}
\end{eqnarray}
Combining these equations with the normalization condition $g_{\mu\nu}u^\mu u^\nu=-1$ and solving for $dr/d\tau$ give
\begin{equation}
\left(\frac{dr}{d\tau}\right)^2=\left({\cal E}-\frac{qQ}{mr}\right)^2-f\left[r^2\left(\frac{d\theta}{d\tau}\right)^2+\frac{\ell^2}{r^2\sin^2{\theta}}+1\right].
\end{equation}
In the equatorial plane where $\theta=\pi/2$, the equation becomes
\begin{equation}
\left(\frac{dr}{d\tau}\right)^2=\left({\cal E}-\frac{qQ}{mr}\right)^2-f\left(\frac{\ell^2}{r^2}+1\right). \label{dr}
\end{equation}
Let us rewrite the last equation in a dimensionless form. We first introduce the following dimensionless quantities:
\begin{equation}
{\cal T}=\frac{\tau}{r_S}, \hspace{1cm} \rho=\frac{r}{r_S}, \hspace{1cm} {\cal L}=\frac{\ell}{r_S} . 
\end{equation}  
Equation~\ref{dr} then becomes
\begin{equation}
\left(\frac{d\rho}{d{\cal T}}\right)^2=\left({\cal E}-\frac{\alpha}{\rho}\right)^2-f\left(\frac{{\cal L}^2}{\rho^2}+1\right), \label{drho}
\end{equation}
where 
\begin{equation}
\alpha=\frac{qQ}{mr_S}.
\end{equation}
The parameter $\alpha$ represent the relative strength of the electromagnetic force to the Newtonian gravitational force. We can rewrite Eq.~\ref{drho} as
\begin{equation}
\left(\frac{d\rho}{d{\cal T}}\right)^2=({\cal E}-V_+)({\cal E}-V_-), \label{drho2}
\end{equation}
where 
\begin{equation}
V_{\pm}=\frac{\alpha}{\rho}\pm \sqrt{f\left(\frac{{\cal L}^2}{\rho^2}+1\right)},
\end{equation}
 is an effective potential. It is $V_+$ that corresponds to physical, future-directed motion and hence will be used in all of the analyses below. Without loss of generality, we will consider ${\cal L}>0$ only.
 
\noindent It was estimated in Ref.~\cite{Zaj2} that the charge of Sgr A* is $10^8-10^{15}$ coulomb. Using the lower limit of charge, the coupling constant for electrons $\alpha_e$ and protons $\alpha_p$ near Sgr A*, which has a mass of $M = 4.3 \times 10^6 M_\odot$ according to Ref.~\cite{GC}, are 
\begin{eqnarray}
\alpha_e \sim 10^9, \\
\alpha_p \sim 10^6.
\end{eqnarray}

\section{Capture Cross-Section of Neutral Particles}\label{S3}
Before we tackle the main problem, let us find the capture cross-cross section for neutral particles first. Setting $\alpha = 0$, the effective potential $V_+$ reduces to
\begin{equation}
V_{+}= \sqrt{f\left(\frac{{\cal L}^2}{\rho^2}+1\right)}.
\end{equation}
%Figure~\ref{fig:Vneut} shows the generic form of $V_+$. 
Capture occurs whenever the particle's energy is greater than the maximum of $V_+$. The function $V_+$ is at an extremum when $dV_+/d\rho=0$ or
\begin{equation}
\rho^2+(3 - 2 \rho)\cl^2=0,\end{equation}
which gives the position of the extrema in terms of $\cl$ as
\begin{equation}
\rho_{\pm} = \cl^2\pm \cl \sqrt{\cl^2-3},
\end{equation}
where $\cl \in [\sqrt{3},\infty)$. When $\cl = \sqrt{3} $ ($\equiv \cl_\text{min}$), $\rho_+$ and $\rho_-$ meet at a saddle point. Inspecting $d^2V_+/d\rho^2$ reveals that $\rho_-$ corresponds to the position of the local maximum of $V_+$. In terms of $\cl$, the escape condition ${\cal E}=V_+|_{\rho=\rho_-}$ becomes
\begin{equation}
{\cal E}=\sqrt{\frac{2}{27}} \left[\cl \left(\sqrt{\cl^2-3}+\cl\right)-\frac{3 \sqrt{\cl^2-3}}{\cl}+9\right]^{1/2},
\end{equation}
Inverting this equation gives
\begin{eqnarray}
\cl = \left[\frac{27 \ce^4-36 \ce^2+\left(9 \ce^2-8\right)^{3/2} \ce+8}{8 \left(\ce^2-1\right)} \right]^{1/2}
\end{eqnarray}
The impact parameter $b$ is defined as the perpendicular distance between the center of force and the incident velocity~\cite{Gold}. It can be written as
\begin{equation}
b = \frac{\cl}{\cal P} = \frac{\cl}{\sqrt{\ce^2-1}},
\end{equation} 
where $\cal P$ is the specific linear momentum. The maximum impact parameter for capture $b_\text{max}$ is given by
\begin{equation}
b_\text{max} = \frac{\left[27 \ce^4-36 \ce^2+\left(9 \ce^2-8\right)^{3/2} \ce+8\right]}{2\sqrt{2} \left(\ce^2-1\right)} ^{1/2}.
\end{equation}
The capture cross-section ${\sigma}_{\text{cap}}$ is given by
\begin{equation}
{\sigma}_{\text{cap}} = \pi b_\text{max}^2 =\frac{\pi}{8} \frac{27 \ce^4-36 \ce^2+\left(9 \ce^2-8\right)^{3/2} \ce+8}{ \left(\ce^2-1\right)^2}.
\end{equation}
 
\begin{figure}
  \centering
  \includegraphics[width=0.7\textwidth]{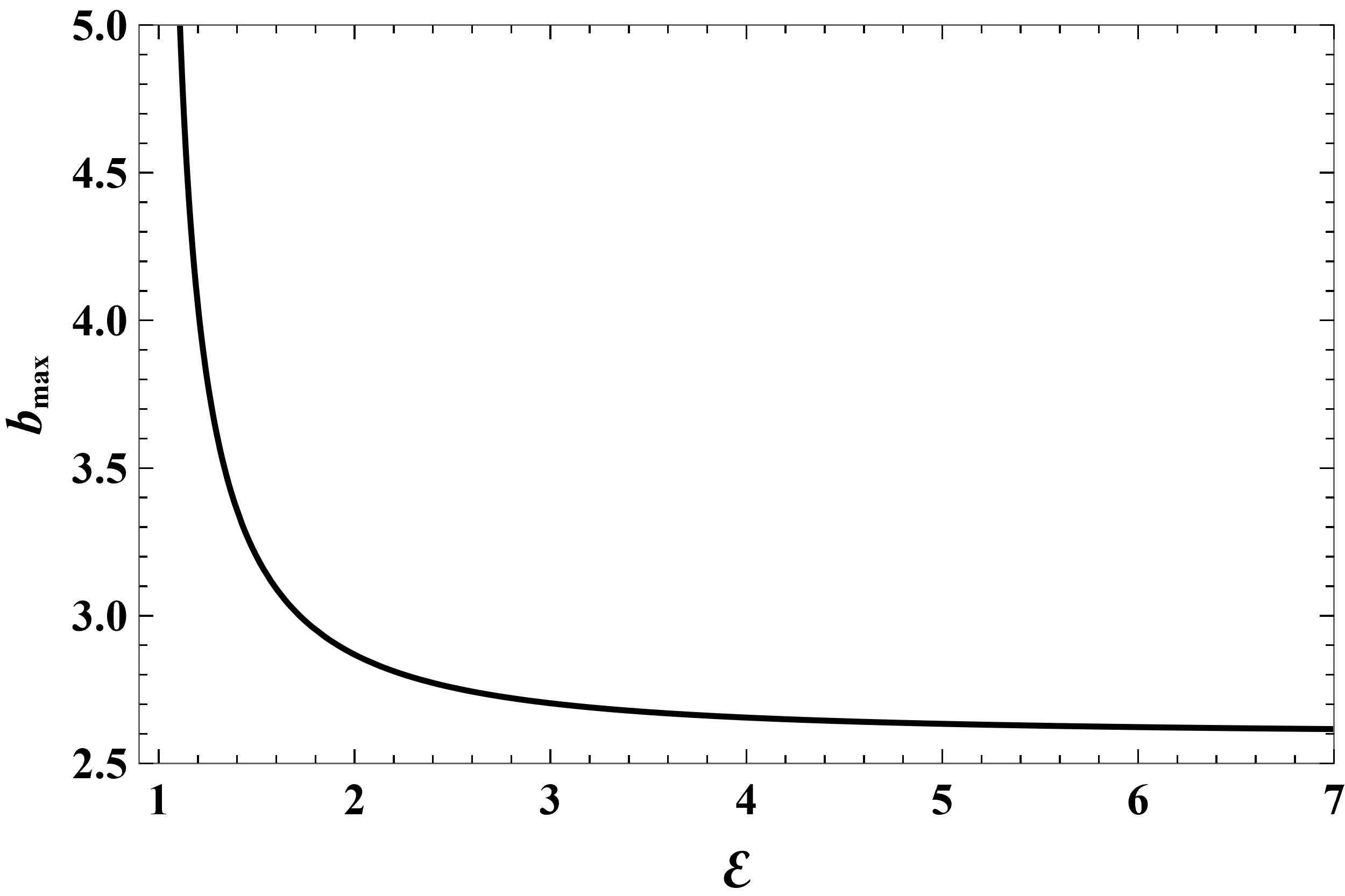}
  \caption{The maximum impact parameter for capture $b_{\text{max}}$ vs. $\ce$ for a neutral particle.}
  \label{fig:bnvsE}
\end{figure}
\begin{figure}
  \centering
  \includegraphics[width=0.7\textwidth]{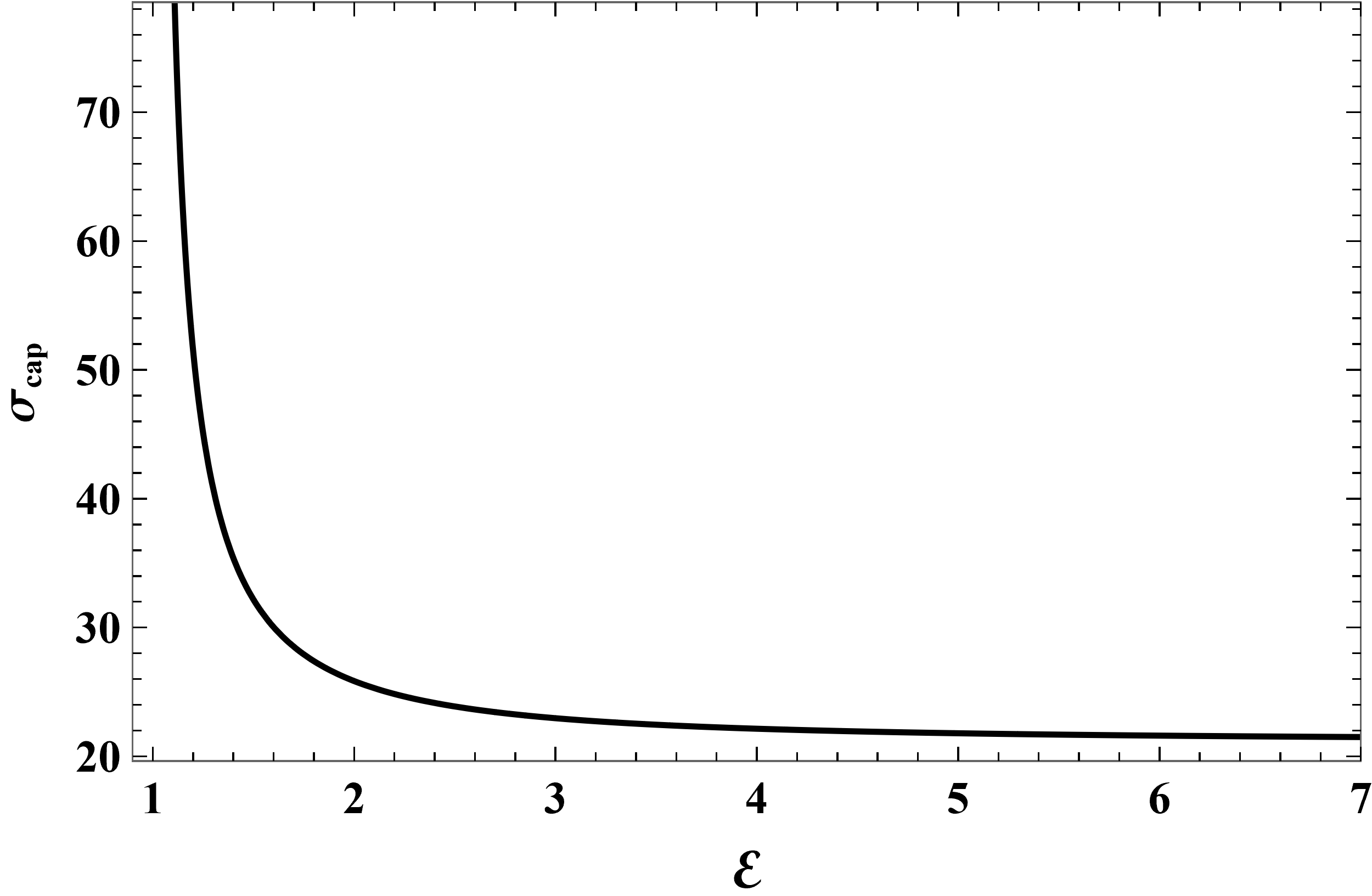}
  \caption{The capture cross-section ${\sigma}_{\text{cap}}$ vs. $\ce$ for a neutral particle.}
  \label{fig:snvsE}
\end{figure}

\noindent Figures~\ref{fig:bnvsE}~and~\ref{fig:snvsE} are plots of $b_{\text{max}}$ and the capture cross-section $\sigma_{\text{cap}}$ vs. $\ce$, respectively. For ultra-relativistic particles (${\cal E}\gg 1$),
\begin{equation}
b_\text{max} = \frac{3 \sqrt{3}}{2}+\frac{\sqrt{3}}{2 {\cal E}^2}+{\mathcal O}\left(\frac{1}{\ce^3}\right).
\end{equation}
The corresponding capture cross-section $\sigma_{\text{cap}}$ is therefore
\begin{equation}
\sigma_{\text{cap}} = \frac{27\pi}{4}+\frac{9\pi}{2{\cal E}^2}+{\mathcal O}\left(\frac{1}{\ce^3}\right).
\end{equation}
For a slowly moving particle with speed $v \ll 1$,
\begin{equation}
\ce\approx 1+\frac{v^2}{2},
\end{equation}
and thus
\begin{equation}
b_\text{max} = \frac{\sqrt{2}}{\sqrt{\ce-1}}+{\mathcal O}(\sqrt{\ce-1}) = \frac{2}{v}+{\mathcal O}(v),
\end{equation}
and the capture cross-section becomes
\begin{equation}
\sigma_{\text{cap}} = \frac{4\pi}{v^2}+{\mathcal O}(v^0).\end{equation}   

\section{Capture Cross-Section of Charged Particles}\label{S4}

\begin{figure}
  \centering
  \includegraphics[width=.7\textwidth]{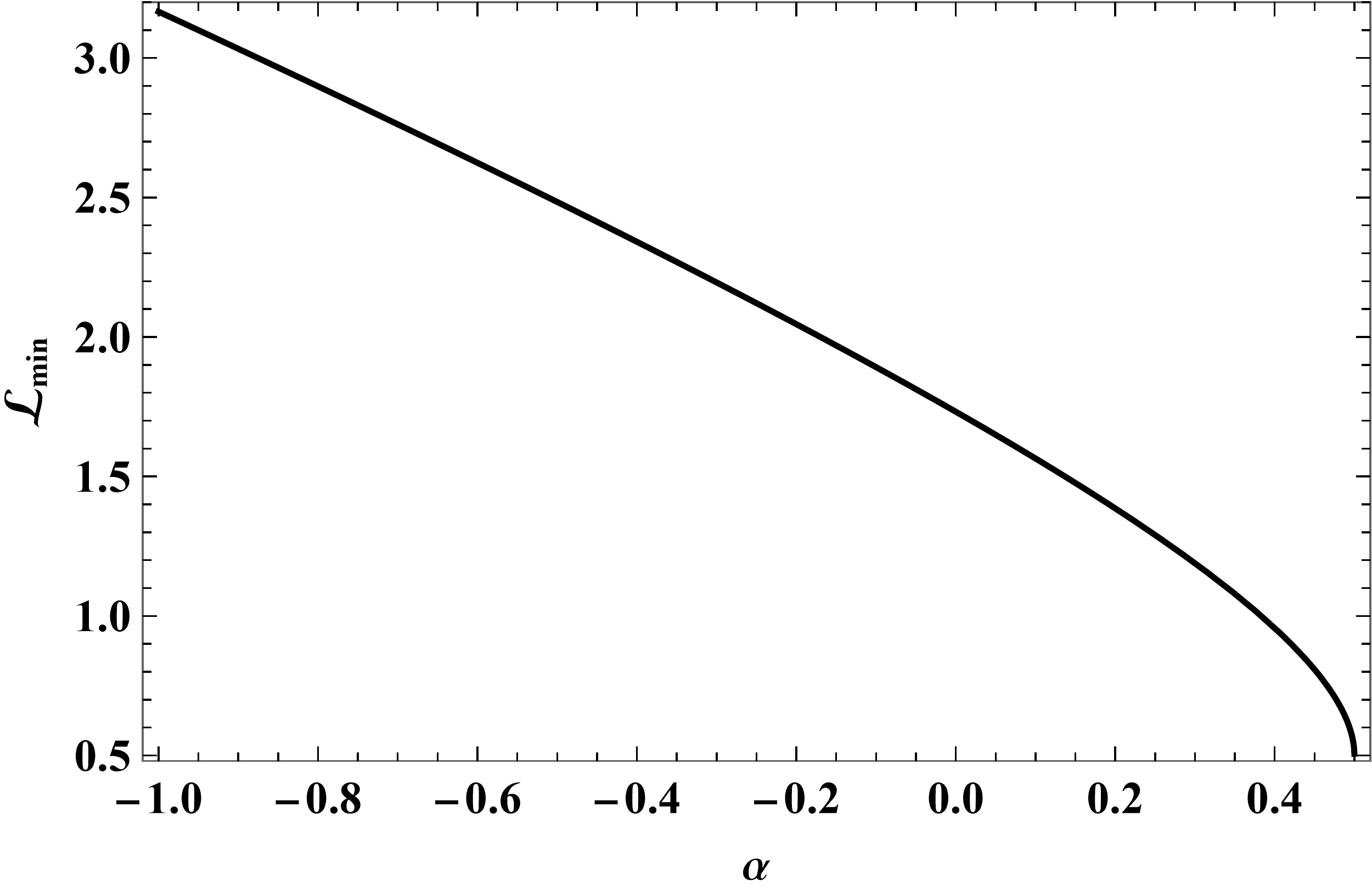}
  \caption{The value of $\cl$ at which $\rho_+$ and $\rho_-$ meet ($\cl_{\text{min}}$)  vs. the electromagnetic coupling parameter $\alpha$.}
  \label{fig:Lmin}
\end{figure}

We will now follow the same procedure we used for the neutral particle. However, analytic expressions are not viable in this case and we will resort to numerical solutions, except in the ultra-relativistic particle case. The structure of the effective potential $V_+$ is generically similar to the neutral particle's. The effect of $\alpha$ is to raise (lower) the peak of $V_+$ for positive (negative) $\alpha$. The effective potential $V_+$ is at an extremum when
\begin{equation}
2 \alpha \sqrt{(\rho_\pm-1) \left(\cl^2+\rho_\pm^2\right) \rho_\pm}-\cl^2 (3-2 \rho_\pm)-\rho_\pm^2 = 0.
\end{equation}
The extremum is a maximum when
\begin{equation}
\frac{\alpha \left[\cl^2 (1-2 \rho_\pm)+(3-4 \rho_\pm) \rho_\pm^2\right]}{\sqrt{(\rho_\pm-1) \left(\cl^2+\rho_\pm^2\right) \rho_\pm}} +2 \rho_\pm - 2 \cl^2 < 0.
\end{equation}
To be consistent with the notation of the previous section, we let $\rho_+$ correspond to the minimum of $V_+$ and $\rho_-$ correspond to the maximum. Here, $\cl_{\text{min}}$ (the value at which $\rho_-$ and $\rho_+$ meet) depends on the value of $\alpha$. The two parameters are related by the relation
\begin{equation}
-\alpha ^8+6 \alpha ^4 \cl_{\text{min}}^2 \left(\cl_{\text{min}}^2-3\right)-8 \alpha ^2 \cl_{\text{min}}^4 \left(\cl_{\text{min}}^2+9\right)+3 \cl_{\text{min}}^4 \left(\cl_{\text{min}}^2-3\right)^2=0.
\end{equation}
\noindent Figure~\ref{fig:Lmin} is a plot of $\cl_{\text{min}}$ vs $\alpha$. When $\alpha = 1/2$, $\cl_{\text{min}}$ approaches zero. This is because $V_+$ ceases to have a local minimum for $\alpha \geq 1/2$. Physically, this limit corresponds to the case when the Coulomb repulsion becomes too strong for stable orbits to exist as discussed in Ref.~\cite{Z1}. 

\begin{figure}
  \centering
  \includegraphics[width=0.7\textwidth]{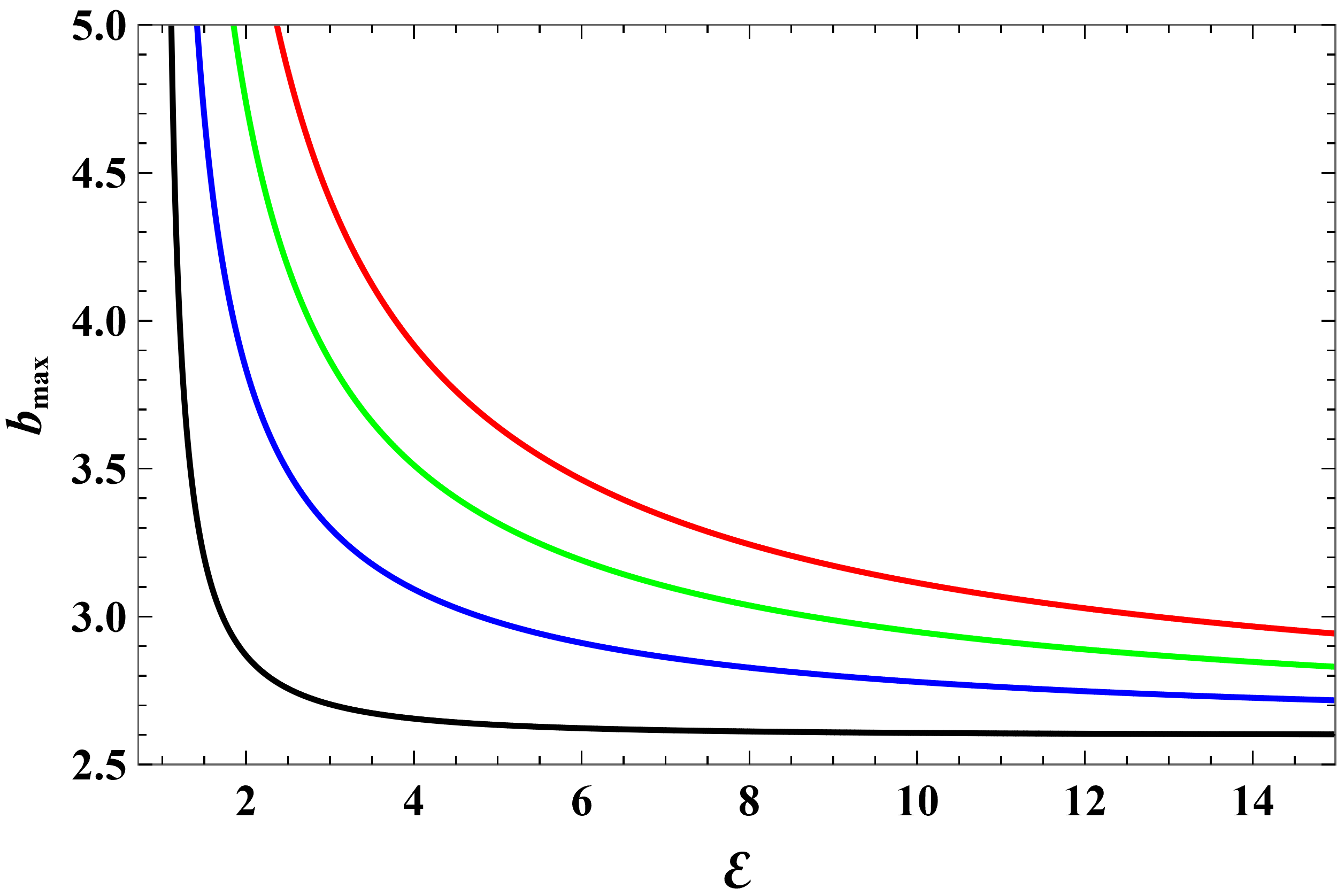}
  \caption{The maximum impact parameter for capture $b_{\text{max}}$ vs. $\ce$ for a charged particle with $\alpha=0$ (black), $\alpha=-1$ (blue), $\alpha=-2$ (green), $\alpha=-3$ (red).}
  \label{fig:bvsE n}
\end{figure}

\noindent Figure~\ref{fig:bvsE n} shows how $b_{\text{max}}$ depends on $\ce$ for several negative values of the coupling parameter $\alpha$. The effect of increasing $|\alpha|$ is to increase the values of $b_{\text{max}}$ for all energies. This is expected because the Coulombs attraction makes it easier for a charges particle to get captured. In all cases, $b_{\text{max}}$ is a monotonic function of $\ce$. In the ultra-relativistic limit, $b_{\text{max}}$ approaches $3\sqrt{3}/2$, the limit in the neutral particle case, for any finite value of $\alpha$, provided that  $\alpha$ is not too large compared to $\ce$.

\begin{figure}
  \centering
  \includegraphics[width=0.7\textwidth]{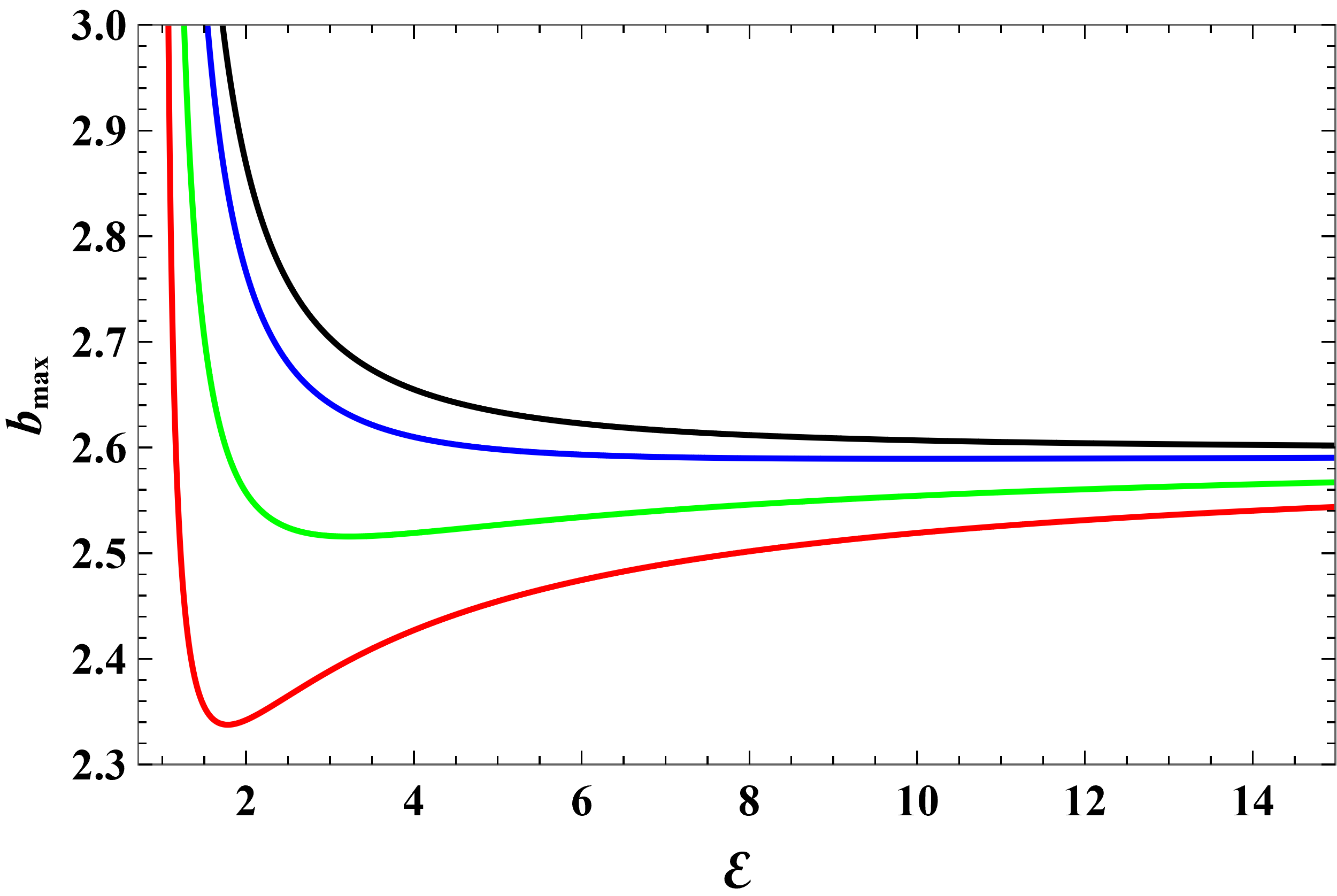}
  \caption{The maximum impact parameter for capture $b_{\text{max}}$ vs. $\ce$ for a charged particle with $\alpha=0$ (black), $\alpha=0.1$ (blue), $\alpha=0.3$ (green), $\alpha=0.5$ (red).}
  \label{fig:bvsE p1}
\end{figure}

\begin{figure}
  \centering
  \includegraphics[width=0.7\textwidth]{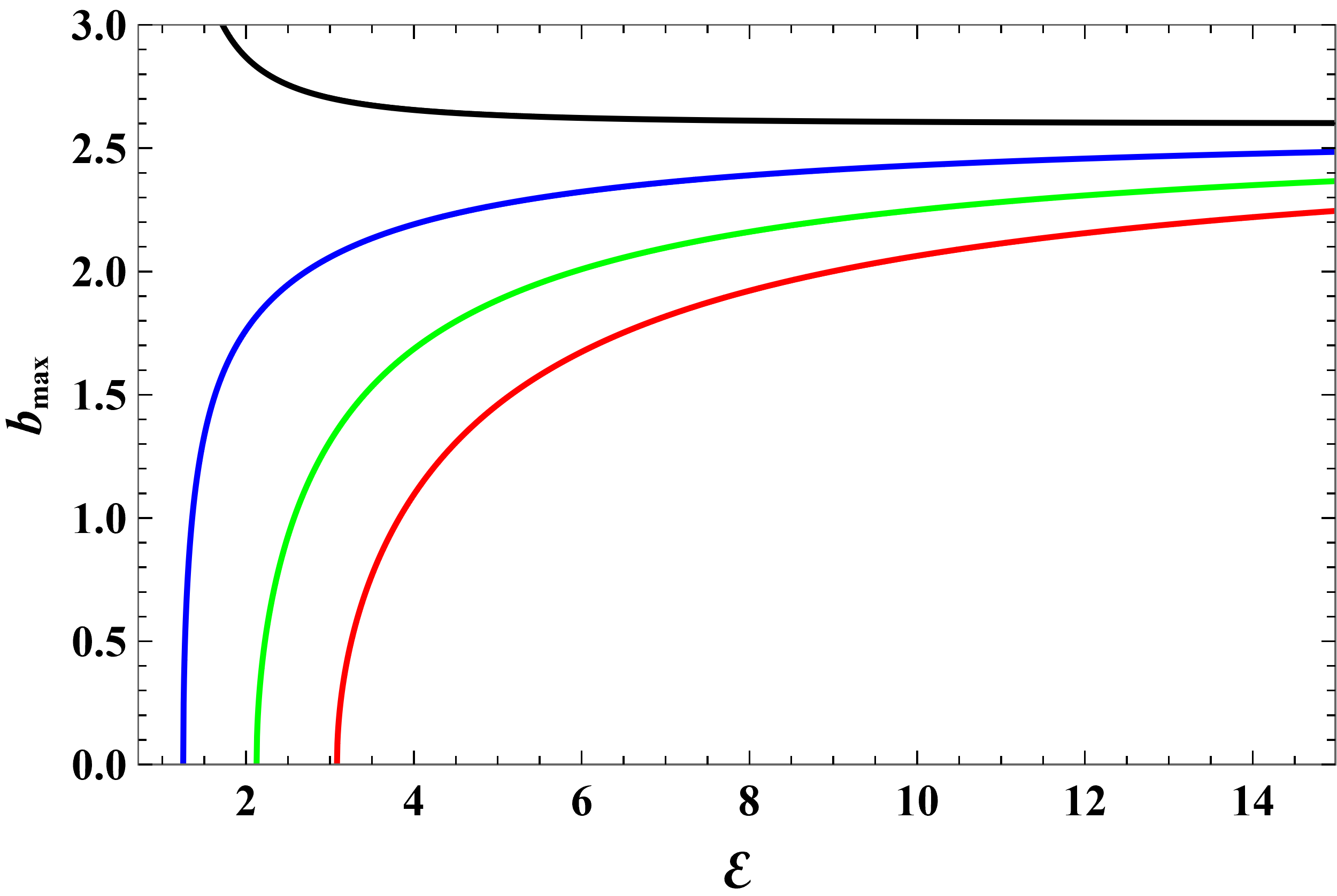}
  \caption{The max impact parameter for capture $b_{\text{max}}$ vs. $\ce$ for a charged particle with with $\alpha=0$ (black), $\alpha=1$ (blue), $\alpha=2$ (green), $\alpha=3$ (red).}
  \label{fig:bvsE p}
\end{figure}

\begin{figure}
  \centering
  \includegraphics[width=.7\textwidth]{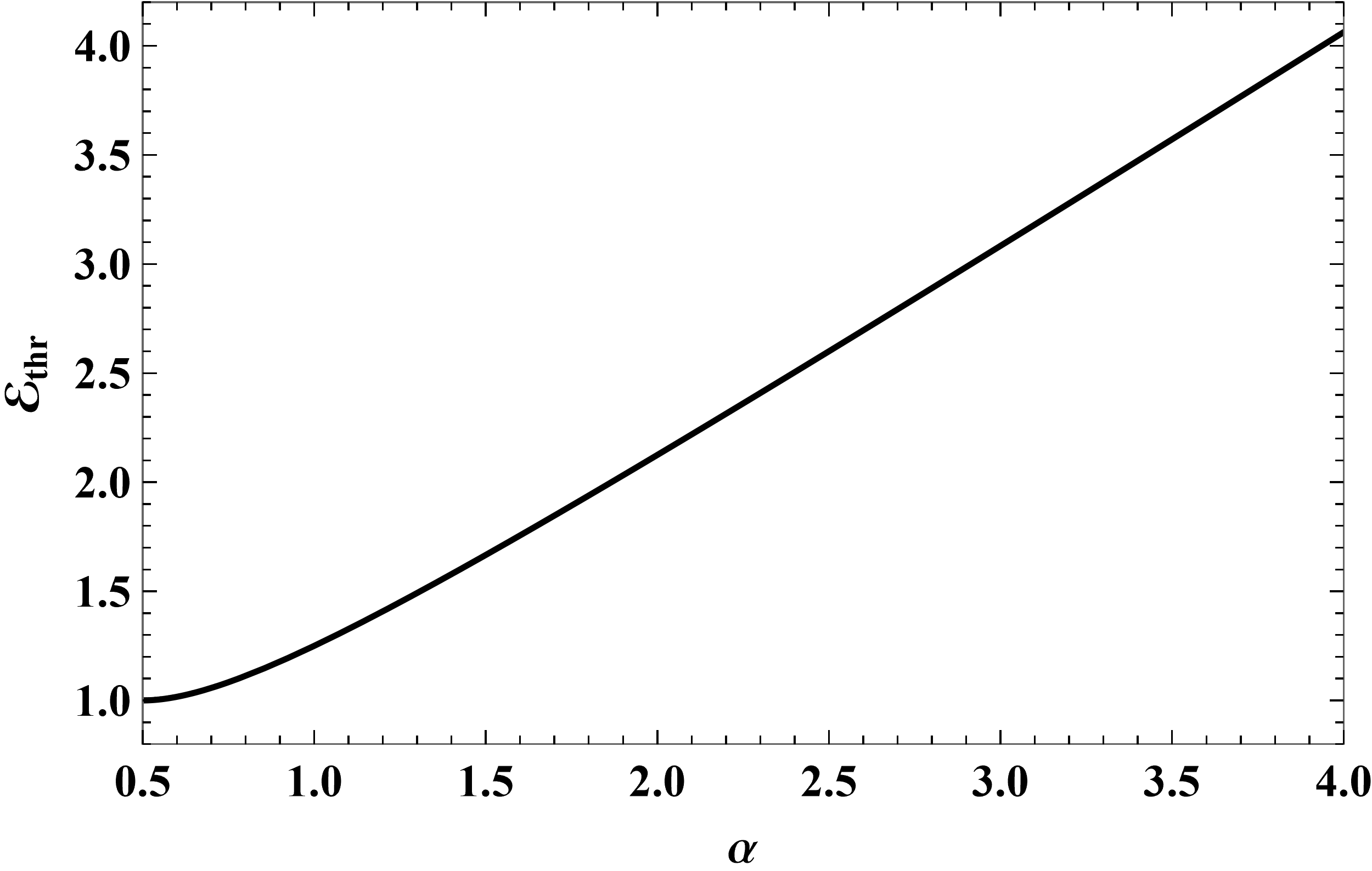}
  \caption{The energy threshold for escape $\ce_{\text{thr}}$  vs. electromagnetic coupling parameter $\alpha$.}
  \label{fig:Ethrvsa}
\end{figure}

\begin{figure}
  \centering
  \includegraphics[width=0.7\textwidth]{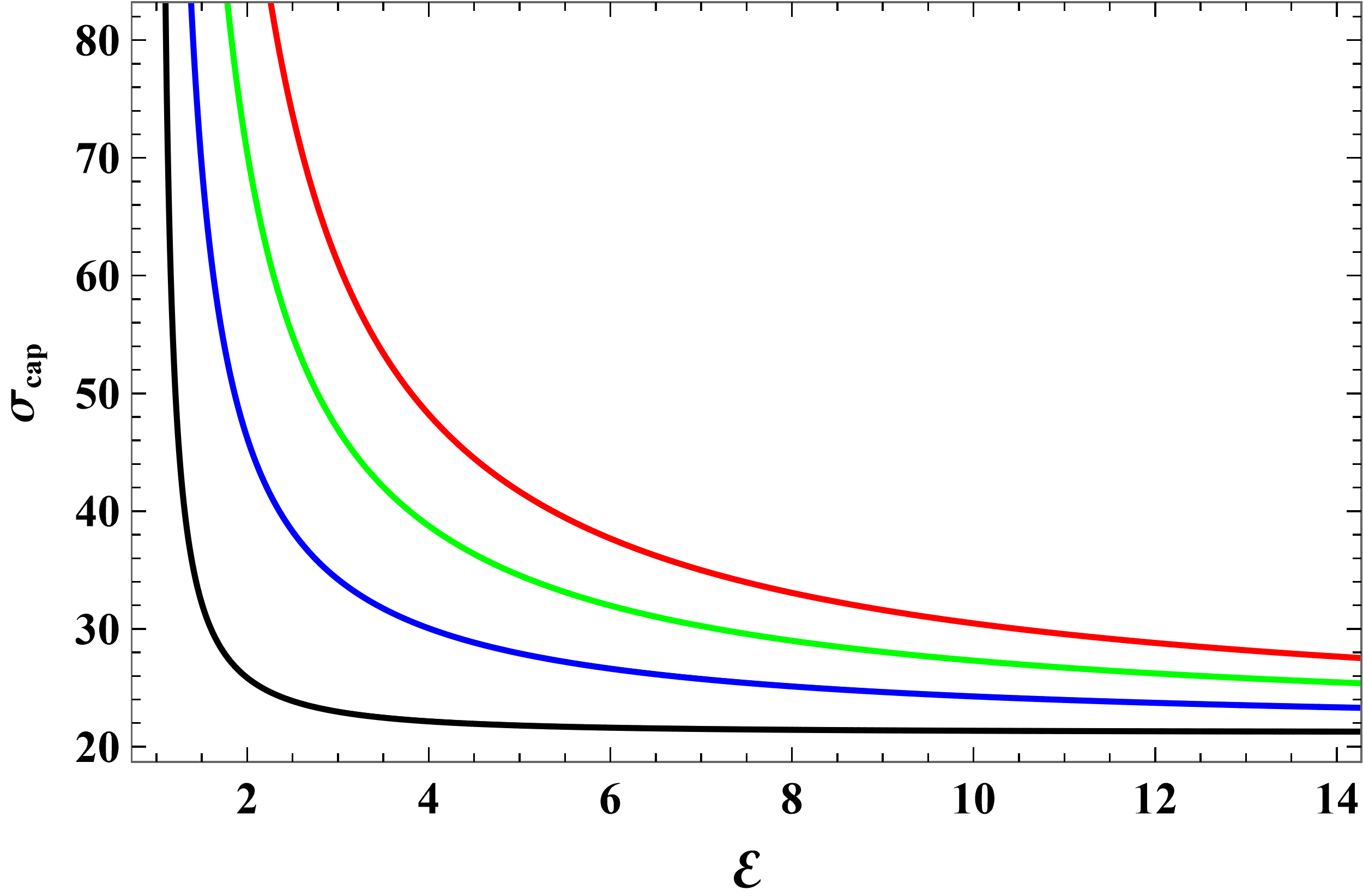}
  \caption{The capture cross-section $\sigma_{\text{cap}}$ vs. $\ce$ for a charged particle with $\alpha=0$ (black), $\alpha=-1$ (blue), $\alpha=-2$ (green), $\alpha=-3$ (red).}
  \label{fig:svsE n}
\end{figure}

\begin{figure}
  \centering
  \includegraphics[width=0.7\textwidth]{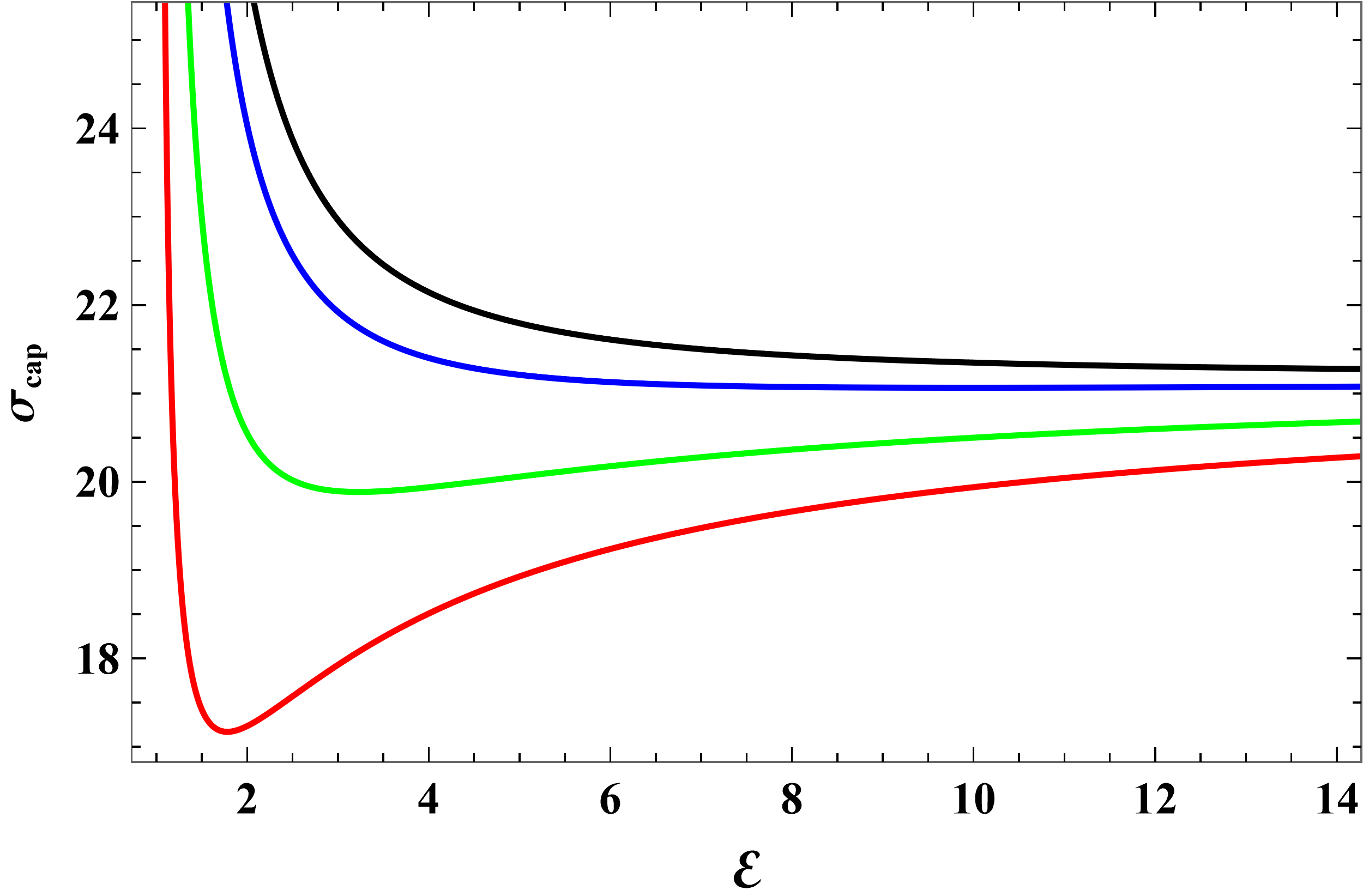}
  \caption{The capture cross-section $\sigma_{\text{cap}}$ vs. $\ce$ for a charged particle with $\alpha=0$ (black), $\alpha=0.1$ (blue), $\alpha=0.3$ (green), $\alpha=0.5$ (red).}
  \label{fig:svsE p1}
\end{figure}

\begin{figure}
  \centering
  \includegraphics[width=0.7\textwidth]{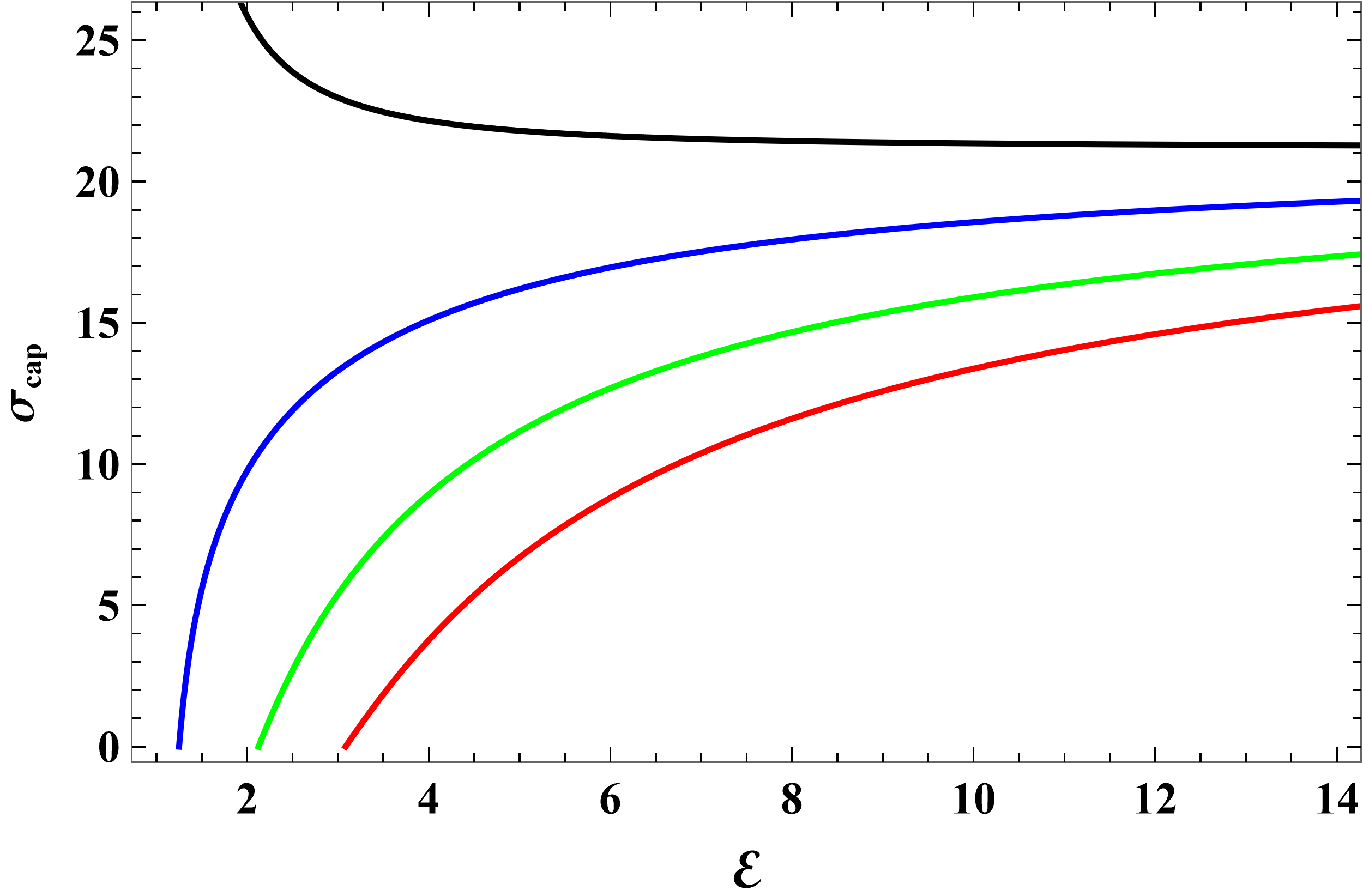}
  \caption{The capture cross-section $\sigma_{\text{cap}}$ vs. $\ce$ for a charged particle with $\alpha=0$ (black), $\alpha=1$ (blue), $\alpha=2$ (green), $\alpha=3$ (red).}
  \label{fig:svsE p}
\end{figure}

\noindent Figure~\ref{fig:bvsE p1} shows how $b_{\text{max}}$ depends on $\ce$ for several values of $\alpha$ between $0$ and $0.5$. In this range, there is competition between that gravitational 'attraction' and the Coulomb repulsion. The curves have richer structure. They falls quickly as $\ce$ goes beyond $1$ and reach a minimum. After that, the curves rise and reach $3\sqrt{3}/2$ asymptotically.
Figure~\ref{fig:bvsE p} shows how $b_{\text{max}}$ depends on $\ce$ for several positive values of $\alpha$ greater than $0.5$. Generally, $b_{\text{max}}$ becomes smaller as $\alpha$ increases. This is expected because the greater the Coulomb repulsion the more difficult it is for a charged particle to be captured. In fact, there is a threshold energy $\ce_\text{thr}$ below which capture cannot occur. It is given by
\begin{equation}
\ce_\text{thr} = \alpha + \frac{1}{4\alpha}.
\end{equation}
This equation is valid for $\alpha \geq 0.5$ only. Fig.~\ref{fig:Ethrvsa} shows how $\ce_\text{thr}$ vary with $\alpha$. 
\\~\\
\noindent The capture cross-section $\sigma_{\text{cap}}$ corresponding to Figs.~\ref{fig:bvsE n}, \ref{fig:bvsE p1} and \ref{fig:bvsE p} is shown in Figs.~\ref{fig:svsE n}, \ref{fig:svsE p} and \ref{fig:svsE p1}, respectively. In all cases, $\sigma_{\text{cap}}$ vs. $\ce$ curves inherent the features of the $b_{\text{min}}$ vs. $\ce$ curves. 


\noindent For ultra-relativistic particles, we can write $b_{\text{max}}$ as 
\begin{equation}
b_{\text{max}} = \frac{3 \sqrt{3}}{2}-\frac{\sqrt{3}\alpha}{\ce}+\frac{9-2\alpha^2}{6\sqrt{3}\ce^2}+{\mathcal O}\left(\frac{1}{\ce^3}\right).
\end{equation}
\noindent The corresponding capture cross-section is then
\begin{equation}
\sigma_{\text{cap}} = \frac{27\pi}{4}-\frac{9\pi\alpha}{\ce}+\frac{\left(4\alpha^2+9\right)\pi}{2\ce^2}+{\mathcal O}\left(\frac{1}{\ce^3}\right).
\end{equation}
These limiting results are in agreement with our numerical findings.

\section{Conclusion}\label{S5}

We have studied the capture cross-section of charged particles by a weakly charged Schwarzschild black hole. We have shown that a trace charge on the black hole can have prominent effects.

When the Coulomb force between a charged particle and the black hole is attractive, it enlarges the capture cross-section significantly. This is expected since the Coulomb attraction enhances the capture of charged particles. However, when the Coulomb force between a charged particle and the black hole is repulsive, it shrinks the capture cross-section significantly. When the electromagnetic coupling strength is below a critical value, capture is possible for all values of the particle's energy. When the electromagnetic coupling strength is above the critical value, there is a minimum value of the particle's energy below which capture is impossible. This is because the Coulomb repulsion surpasses the gravitational attraction unless the particle's radial momentum is large enough.  

Our results emphasizes the assertion that charged black holes will favorably accretes charges of the opposite sign. However, it is still possible for the black hole charge to grow if the plunging charged particles are energetic enough to the limit that the capture cross-section becomes independent of the sign of the charges. Moreover, the fact that the electromagnetic coupling constant is three orders of magnitudes greater for electrons than protons suggests that it is relatively easier for a black hole to accumulate positive charge than negative charge.

It will be an astrophysically interesting to study the energies of charged particles near an astrophysical black hole to understand better how the black hole's charge evolves. The problem can be astrophyically more viable when other astrophysical black holes, such as rotating black holes, are studied (in progress).

\label{lastpage}

\end{document}